\documentclass[12pt]{article}
\usepackage{amssymb}

\input{tcilatex}

\begin{document}

\smallskip\ 

\begin{center}
\textbf{TOWARD A CONNECTION BETWEEN THE ORIENTED}

\smallskip\ 

\textbf{MATROID THEORY AND SUPERSYMMETRY}

\textbf{\ }

\smallskip\ 

J. A. Nieto \footnote{%
nieto@uas.uasnet.mx}

\smallskip

\textit{Facultad de Ciencias F\'{\i}sico-Matem\'{a}ticas, Universidad Aut%
\'{o}noma}

\textit{de Sinaloa, C.P. 80000, Culiac\'{a}n Sinaloa, M\'{e}xico}

\bigskip\ 

\bigskip\ 

\textbf{Abstract}
\end{center}

We considered the possibility that the oriented matroid theory is connected
with supersymmetry via the Grassmann-Plucker relations. The main reason for
this, is that such relations arise in both in the chirotopes definition of
an oriented matroid, and in maximally supersymmetric solutions of eleven-
and ten-dimensional supergravity theories. Taking this observation as a
motivation, and using the concept of a phirotope, we propose a mechanism to
implement supersymmetry in the context of the oriented matroid theory.

\bigskip\ 

\bigskip\ 

\bigskip\ 

\bigskip\ 

\bigskip\ 

\bigskip\ 

\bigskip\ 

Keywords: oriented matroid theory, maximal supersymmetry in supergravity,
chirotopes

Pacs numbers: 04.60.-m, 04.65.+e, 11.15.-q, 11.30.Ly

October, 2005

\newpage \noindent \textbf{1.- Introduction}

\smallskip\ 

It has been shown [1]-[6] through the chirotope concept that oriented
matroid theory [7] is related to several sectors of $M$-theory, including
Chern-Simons theory, supergravity, string theory, and $p$-branes physics.
These connections motivated a recent proposal [8] of considering the
oriented matroid theory as the underlying mathematical framework for $M$%
-theory. But due the expected fermionic structure of $M$-theory is almost
impossible to avoid wondering about supersymmetry in this scenario. In fact,
until now all connections between matroids and $M$-theory have been realized
in the bosonic sector. The main reason for this development has to do with a
technical reason; as far as we know, mathematically oriented matroid theory
has not been linked to supersymmetry.

In this work, we argue that the phirotope concept [9] which is a
generalization of the chirotope concept (see Ref. [7]) may provide the route
for such a link. As it is known, the chirotope concept determines one
possible definition of an oriented matroid. In the case of realizable
oriented matroid the chirotope can be defined in terms of the
Grassmann-Plucker relations over the real ones. It turns out that these
ideas can be transferred from real to complex structure and in this case the
chirotope is called phirotope [9] (see also Refs. [10] and [11]). As we
shall explain, the phirotope is an alternating function over the complex one
that satisfies a generalized Grassmann-Plucker relations.

Another source of motivation for the present work arose when we observed
that surprisingly, the Grassmann-Plucker relations have played a very
important role in a related subject: the classification of maximally
supersymmetric solutions of ten- and eleven-dimensional supergravity
theories. In fact, Figueroa-O'Farrill and Papadopoulos [12]-[13] had shown
that maximal supersymmetry in eleven-dimensional (ten-dimensional)
supergravity leads to a quadratic condition for the four-form $F$ (five-form 
$F$) which implies the Grassmann-Plucker relations for $F$. From this result
they conclude that the AdS solutions, the Hpp-waves, and flat space
solutions exhaust the maximally supersymmetric solutions of
eleven-dimensional supergravity. Similar conclusions hold in the case of
ten-dimensional supergravity.

Here, we claim that looking at the $F$ field as a chirotope the
Figueroa-O'Farrill-Papadopoulos' construction can be understood as a link
between the oriented matroid theory and the maximal supersymmetry in ten-
and eleven-dimensional supergravities.

Just for introducing some notation and definitions in section 2 we briefly
review the notions of chirotope and phirotope. In section 3, we summarize
the maximally supersymmetric solutions of eleven-dimensional supergravity
constrains for the $F$ field and compare them with the chirotope concept via
the Grassmann-Plucker relations. In section 4, we present a possible
definition of a superphirotope. Finally, in section 5, we make some last
comments.

\bigskip\ 

\noindent \textbf{2.- Chirotope and Phirotope concepts}

\smallskip\ 

Let us start introducing the completely antisymmetric symbol

\begin{equation}
\varepsilon ^{a_{1}...a_{d}}\in \{-1,0,1\}.  \tag{1}
\end{equation}%
Here the indices $a_{1},...,a_{d}$ run from $1$ to $d$. This is a $d$-rank
tensor which values are $+1$ or $-1$ depending on even or odd permutations
of $\varepsilon ^{12...d},$ respectively. Moreover, $\varepsilon
^{a_{1}...d_{d}}$ takes the value $0$ unless $a_{1}...a_{d}$ are all
different. Let $v_{a}^{i}$ be any $d\times n$ matrix over some field $F,$
where the index $i$ takes values in the set $E=\{1,...,n\}$. Consider the
object%
\begin{equation}
\Sigma ^{i_{1}...i_{d}}=\varepsilon ^{a_{1}...a_{d}}v_{a_{1}}^{i_{1}}\cdot
\cdot \cdot v_{a_{d}}^{i_{d}},  \tag{2}
\end{equation}%
which can also be written as%
\begin{equation}
\Sigma ^{i_{1}...i_{d}}=\det (\mathbf{v}^{i_{1}},...,\mathbf{v}^{i_{d}}). 
\tag{3}
\end{equation}%
Using the $\varepsilon -$symbol property

\begin{equation}
\varepsilon ^{a_{1}...[a_{d}}\varepsilon ^{b_{1}...b_{d}]}=0.  \tag{4}
\end{equation}%
It is not difficult to prove that $\Sigma ^{i_{1}...i_{d}}$ satisfies the
so-called Grassmann-Plucker relations, namely

\begin{equation}
\Sigma ^{i_{1}...[i_{d}}\Sigma ^{j_{1}...j_{d}]}=0.  \tag{5}
\end{equation}%
The brackets in the indices of (4) and (5) mean completely antisymmetrized.

A realizable chirotope $\chi $ is defined as

\begin{equation}
\chi ^{i_{1}...i_{d}}=sign\Sigma ^{i_{1}...i_{d}}.  \tag{6}
\end{equation}%
From the point of view of exterior algebra one finds that there is a close
connection between Grassmann algebra and a chirotope. Let us denote by $%
\wedge _{d}R^{n}$ the $(_{d}^{n})$-dimensional real vector space of
alternating $d$-forms on $R^{n}$. An element $\mathbf{\Sigma }$ in $\wedge
_{d}R^{n}$ is said to be decomposable if

\begin{equation}
\mathbf{\Sigma }=\mathbf{v}_{1}\wedge \mathbf{v}_{2}\wedge ...\wedge \mathbf{%
v}_{d},  \tag{7}
\end{equation}%
for some $\mathbf{v}_{1},\mathbf{v}_{2},...,.\mathbf{v}_{d}\in R^{n}$. It is
not difficult to see that (7) can be written as

\begin{equation}
\mathbf{\Sigma }=\frac{1}{r!}\Sigma ^{i_{1}...i_{d}}e_{i_{1}}\wedge
e_{i_{2}}\wedge ...\wedge e_{i_{d}},  \tag{8}
\end{equation}%
where $e_{i_{1}},e_{i_{2}},...,e_{i_{d}}$ are $1$-form bases in $R^{n}$ and $%
\Sigma ^{i_{1}...i_{d}}$ is given in (3). This shows that $\Sigma
^{i_{1}...i_{d}}$ can be identified with an alternating decomposable $d$%
-form.

In order to define non-realizable chirotopes it is convenient to write the
expression (5) in an alternative form

\begin{equation}
\tsum\limits_{k=1}^{d+1}s_{k}=0,  \tag{9}
\end{equation}%
where

\begin{equation}
s_{k}=(-1)^{k}\Sigma ^{i_{1}...i_{d-1}j_{k}}\Sigma ^{j_{1}...\hat{\jmath}%
_{k}...j_{d+1}}.  \tag{10}
\end{equation}%
Here, $j_{d+1}=i_{d}$ and $\hat{\jmath}_{k}$ establish the notation for
omitting this index. Thus, in general for any $d$-rank chirotope $\chi
:E^{d}\rightarrow \{-1,0,1\}$ and

\begin{equation}
s_{k}=(-1)^{k}\chi ^{i_{1}...i_{d-1}j_{k}}\chi ^{j_{1}...\hat{\jmath}%
_{k}...j_{d+1}},  \tag{11}
\end{equation}%
for $k=1,...,d+1$ there exist $r_{1},...,r_{d+1}\in R^{+}$ such that

\begin{equation}
\tsum\limits_{k=1}^{d+1}r_{k}s_{k}=0.  \tag{12}
\end{equation}%
It is clear that (9) is a particular case of (12). Therefore, there are
chirotopes that may be non-realizable. Moreover, this definition of a
chirotope is equivalent to various others (see Ref. [7] for details), but
the present one is more convenient for a generalization to the complex
structure setting.

The generalization of a chirotope to a phirotope is straightforward. A
function $\varphi :E^{d}\rightarrow S^{1}\cup \{0\}$ on all $d$-tuples of $%
E=\{1,...,n\}$ is called a $d$-rank phirotope if (a) $\varphi $ is
alternating and (b) for

\begin{equation}
\omega _{k}=(-1)^{k}\varphi ^{i_{1}...i_{d-1}j_{k}}\varphi ^{j_{1}...\hat{%
\jmath}_{k}...j_{d+1}}=0,  \tag{13}
\end{equation}%
for for $k=1,...,d+1$ there exist $r_{1},...,r_{d+1}\in R^{+}$ such that

\begin{equation}
\tsum\limits_{k=1}^{d+1}r_{k}\omega _{k}=0.  \tag{14}
\end{equation}

In the case of a realizable phirotope we have

\begin{equation}
\Omega ^{i_{1}...i_{d}}=\omega (\det (\mathbf{u}^{i_{1}},...,\mathbf{u}%
^{i_{d}})),  \tag{15}
\end{equation}%
where $\omega (z)\in S^{1}\cup \{0\}$ and $(\mathbf{u}^{i_{1}}...\mathbf{u}%
^{i_{d}})$ are a set of complex vectors in $C^{d}$. We observe that one of
the main differences between a chirotope and a phirotope is that the image
of a phirotope is no longer a discrete set (see Ref. [9] for details).

\bigskip\ 

\noindent \textbf{3- Maximal supersymmetry in eleven-dimensional
supergravity and the chirotope concept}

\smallskip\ 

In Ref. [12] Figueroa-O'Farril and Papadopoulos showed that maximal
supersymmetry in eleven-dimensional supergravity implies the two conditions

\begin{equation}
F_{M[L_{1}L_{2}L_{3}}F_{L_{4}L_{5}L_{6}L_{7}]}=0  \tag{16}
\end{equation}%
and

\begin{equation}
F_{M[P_{1}P_{2}P_{3}}F_{Q_{1}Q_{2}Q_{3}]N}=0,  \tag{17}
\end{equation}%
for the $4$-form field strength $F=dA$ in eleven dimensions$.$ Moreover,
from (16) and (17) they showed that $F$ is parallel and decomposable. This
last property means that $F$ satisfies the Grassmann-Plucker relations

\begin{equation}
F_{MP_{1}P_{2}[P_{3}}F_{Q_{1}Q_{2}Q_{3}Q_{4}]}=0.  \tag{18}
\end{equation}%
Thus, according to the discussion of the previous section we discover that
(18) establishes that $F$ is a realizable $4$-rank chirotope with a ground
set $E=\{1,...,11\}$. This in turn means that maximal supersymmetry in
eleven-dimensional supergravity is related to oriented matroid theory.
Similar conclusion can be obtained for the case of ten-dimensional
supergravity (see section 5). Hence, from the chirotope concept we have
found a link between supersymmetry and the oriented matroid theory.
Therefore, one should expect a generalization of oriented matroid theory
which would include supersymmetry. But in order to develop this idea it
turns out more convenient to consider a complex structure, and this means
that we need to focus on the superphirotope notion rather than on the
superchirotope concept which should arise as a particular case of the former.

\bigskip\ 

\noindent \textbf{4. Superphirotope}

\smallskip\ 

The main goal of this section is to outline a possible supersymmetrization
of a phirotope. Because of convenience we shall call superphirotope such a
supersymmetric phirotope. Inspired in supebrane theory we find that one way
to define a superphirotope, which assures supersymmetry, is as follows.
First, we need to locally consider the expressions (13)-(15) in the sense
that $\varphi ^{i_{1}....j_{d}}(\xi )$ is a local phirotope if

\begin{equation}
\omega _{k}=(-1)^{k}\varphi ^{i_{1}...i_{d-1}j_{k}}(\xi )\varphi ^{j_{1},...%
\hat{\jmath}_{k}...j_{d+1}}(\xi ),  \tag{19}
\end{equation}%
for $k=1,...,d+1$ there exist $r_{1},...,r_{d+1}\in R^{+}$ such that

\begin{equation}
\tsum\limits_{k=1}^{d+1}r_{k}\omega _{k}(\xi )=0.  \tag{20}
\end{equation}

In the case of a realizable local phirotope we have

\begin{equation}
\Omega ^{i_{1}...i_{d}}(\xi )=\omega (\det (\mathbf{u}^{i_{1}}(\xi ),...,%
\mathbf{u}^{i_{d}}(\xi )),  \tag{21}
\end{equation}%
where $\xi =(\xi ^{1},...,\xi ^{d})$ are local coordinates of some $d-$%
dimensional manifold $B$. The vectors $\mathbf{v}^{i_{1}}(\xi ),...,\mathbf{v%
}^{i_{d}}(\xi )$ can be thought as vectors in the tangent space $T_{\xi }(B)$
at $\xi $. One can assume that the possibility of considering the
expressions (19)-(21) in a local context may be justified in principle by
the so-called matroid bundle notion [14]-[17]. It is known that the
projective variety of decomposable forms is isomorphic to the Grassmann
variety of $d$-dimensional linear subspaces in $R^{n}$. In turn, the
Grassmann variety is the classifying space for vector bundle structures.
Taking these ideas as a motivation, MacPherson [14] developed the
combinatorial differential manifold concept. The matroid bundle concept
[15]-[17] arises as a generalization of the MacPherson proposal. Roughly
speaking, a matroid bundle is a structure in which at each point of the
differentiable manifold an oriented matroid is attached as a fiber (see
[14]-[17] for details).

Now, let us consider a supermanifold $\mathcal{B}$ parametrized by the local
coordinates $(\xi ,\theta )$ where $\theta $ are elements of the odd
Grassmann algebra (anticommuting variables). We shall now consider the
supersymmetric prescription

\begin{equation}
\mathbf{v}^{i}\rightarrow \mathbf{\pi }^{i}=\mathbf{v}^{i_{1}}-i\bar{\theta}%
\gamma ^{i}\mathbf{\partial }\theta .  \tag{22}
\end{equation}%
Here, $\gamma ^{i}$ are elements of a Clifford algebra. Using (22) we can
generalize (21) in the form

\begin{equation}
\Psi ^{i_{1}...i_{d}}(\xi ,\theta )=\omega (\det (\mathbf{\pi }^{i_{1}}(\xi
,\theta ),...,\mathbf{\pi }^{i_{d}}(\xi ,\theta )).  \tag{23}
\end{equation}%
Here, the symbol $\det $ means the superdeterminant. One should expect that
(23) satisfies a kind of supersymmetric Grassmann-Plucker relations. It is
not difficult to see that up to total derivative (23) is invariant under the
global supersymmetric transformations

\begin{equation}
\delta \theta =\epsilon  \tag{24}
\end{equation}%
and

\begin{equation}
\delta \mathbf{v}^{i_{1}}=i\bar{\epsilon}\gamma ^{i}\partial \theta , 
\tag{25}
\end{equation}%
where $\epsilon $ is a constant complex spinor parameter.

Similarly, one can generalize the superphirotope to the non-representable
case by assuming that if

\begin{equation}
\omega _{k}=(-1)^{k}\varphi ^{i_{1}...i_{d-1}j_{k}}(\xi ,\theta )\varphi
^{j_{1}...\hat{\jmath}_{k}....j_{d+1}}(\xi ,\theta ),  \tag{26}
\end{equation}%
for $k=1,...,d+1$ there exist $r_{1},...,r_{d+1}\in R^{+}$ such that

\begin{equation}
\tsum\limits_{k=1}^{d+1}r_{k}\omega _{k}(\xi ,\theta )=0.  \tag{27}
\end{equation}%
Of course, in the case that the complex structure is projected to the real
structure one should expect that the superphirotope is reduced to the
superchirotope.

\bigskip\ 

\noindent \textbf{5. Final remarks}

\smallskip\ 

With the superphirotope $\Psi ^{i_{1}...i_{d}}(x,\theta )$ at hand one may
consider a possible partition function%
\begin{equation}
Z=\int D\Psi \exp (iS),  \tag{28}
\end{equation}%
where

\begin{equation}
S=\frac{1}{2}\int d^{d}\xi d\theta (\lambda ^{-1}\Psi ^{i_{1}...i_{d}}(\xi
,\theta )\Psi _{^{i_{1}...i_{d}}}(\xi ,\theta )-\lambda T_{d}^{2})  \tag{29}
\end{equation}%
is a Schild type action for a superphirotope. Here, $\lambda $ is a Lagrange
multiplier and $T_{d}$ is the $(d-1)$-phirotope tension. Moreover, in a more
general context the action may have the form

\begin{equation}
S=\frac{1}{2}\int d^{d}\xi d\theta (\lambda ^{-1}\varphi
^{_{^{i_{1}....i_{d}}}}(\xi ,\theta )\varphi _{^{i_{1}...i_{d}}}(\xi ,\theta
)-\lambda T_{d}^{2}).  \tag{30}
\end{equation}%
The advantage of the actions (29) and (30) is that duality is assured in an
automatic way. In fact, in oriented matroid theory duality is a main subject
in the sense that any chirotope has an associated dual chirotope. This means
that a theory described in the context of an oriented matroid automatically
contents a duality symmetry. Therefore, with our prescription we are
assuring not only the supersymmetry for the action (29) or (30) but also the
duality symmetry.

The action (29) can be related to an ordinary super $p$-brane by assuming
that $\Psi ^{i_{1}...i_{d}}(\xi ,\theta )$ is a closed $d$-form because in
that case we can write

\begin{equation}
\pi _{a}^{i}=\partial _{a}x^{i}-i\bar{\theta}\gamma ^{i}\partial _{a}\theta .
\tag{31}
\end{equation}%
Here, the coordinates $x^{i}$ are the $p$-brane bosonic coordinates. It is
worth mentioning that the closedness of the bosonic sector of $\Psi
^{i_{1}...i_{d}}(\xi ,\theta )$ is a constraint of Nambu-Poisson geometry
which has been related to oriented matroid theory (see Ref. [18] for
details).

It may be interesting for further research to consider the action (29) from
the point of view of a superfield formalism instead of using the
prescription (31). In this case one may consider a supersymmetrization in
the form $\pi _{a}^{i}(\xi ,\theta )=\partial _{a}X^{i}$, with $X^{i}$ as a
scalar superfield admitting a finite expansion in terms of $\theta .$ For
instance, in four dimensions we have

\begin{equation}
X^{i}(\xi ,\theta )=x^{i}(\xi )+i\theta \psi ^{i}(\xi )+\frac{i}{2}\bar{%
\theta}\theta B^{i}(\xi ).  \tag{32}
\end{equation}%
Here, $\psi ^{i}$ is a Majorana spinor field and $B^{i}$ is an auxiliary
field. By substituting (32) into (23) one should expect a splitting of (29)
in several terms containing the variables $x^{i}(\xi ),\psi ^{i}(\xi )$ and $%
B^{i}(\xi )$. The important thing is that using the prescription (32)
supersymmetry becomes evident in the sense that the algebra of supersymmetry
transformations is closed off the mass-shell (see Ref. [19]).

Although in section 3 we focused on eleven-dimensional supergravity similar
arguments can be applied to the case of ten-dimensional supergravity.
Specifically, by studying maximal supersymmetry in IIB supergravity
Figueroa-O'Farril and Papadopoulos [12] used the vanishing of the curvature
of the supercovariant derivative to derive the analogue Grassmann-Plucker
formula

\begin{equation}
F_{LP_{1}P_{2}P_{3}[P_{4}}F_{Q_{1}Q_{2}Q_{3}Q_{4}]}^{L}=0,  \tag{33}
\end{equation}%
for the five-form $F_{LP_{1}P_{2}P_{3}P_{4}}$. Moreover, in Ref. [13] is
proved that (33) implies that

\begin{equation}
F=G+^{\ast }G,  \tag{34}
\end{equation}%
where $G$ is a decomposable five-form and $^{\ast }G$ denotes the
ten-dimensional dual of $G.$ This means that $G$ and $^{\ast }G$ satisfy the
Grassmann-Plucker relations and therefore can be identified with a $5-$rank
chirotope. Thus, we conclude that not only maximal supersymmetry in eleven
dimensional supergravity implies a link between supersymmetry and the
oriented matroid theory via the chirotope concept but also maximal
supersymmetry in ten-dimensional supergravity.

\smallskip\ 

\textbf{Acknowledgment: }I would like to thank R. Alarcon for his
hospitality in the Physics \& Astronomy Department of the Arizona State
University where part of this work was developed. I would also like to thank
G. Hurlbert for his helpful comments.

\end{document}